\documentclass{emulateapj}

\newcommand{\dgr}{$^{\circ}$}
\newcommand{\etal}{et~al.\ }

\slugcomment{To Appear in The Astrophysical Journal}

\shorttitle{SDSS BL~Lac Objects}
\shortauthors{Smith \etal}

\begin{document}

\title{Highly Polarized Optically-Selected BL~Lacertae Objects}
\vskip 0.2in

\author{Paul S. Smith\altaffilmark{1}, G. Grant Williams\altaffilmark{2},
Gary D. Schmidt\altaffilmark{1}, Aleksandar M. Diamond-Stanic\altaffilmark{1},
\& Dennis L. Means\altaffilmark{1}}
\altaffiltext{1}{Steward Observatory, The University of Arizona,
    Tucson, AZ 85721; psmith, schmidt, adiamond, dmeans@as.arizona.edu}

\altaffiltext{2}{MMTO, The University of Arizona, Tucson, AZ 85721; gwilliams@mmto.org}

\begin{abstract}

Observations of candidate BL~Lacertae objects
spectroscopically selected from the Sloan Digital Sky Survey (SDSS)
reveal a large fraction with high polarization ($P > 3$\%).
This result confirms that synchrotron radiation makes an important
contribution to the observed optical continuum for most objects in the sample.
The SDSS sample can be divided into
separate categories, with objects of undetermined redshift generally having
the highest optical polarization.
Polarization as high as 23\% and the lack of spectral features
suggests that the 
synchrotron continuum completely dominates the spectra of these 
sources.
The mean polarization levels observed for objects having measured
redshifts is much lower,
with the maximum observed polarization for this group being
$\sim$10\%.
The lower polarizations of these objects are reminiscent
of the less spectacular
polarization levels shown by BL~Lac objects discovered in X-ray
surveys.
We find no SDSS BL~Lac candidates at $z \gtrsim 1$ with $P > 3$\%, calling their
classification as BL~Lac objects into question.
In addition, the existence of radio-quiet BL~Lac objects is not verified since
none of 10 potentially radio-weak BL~Lac candidates observed
are highly polarized.
Regardless of whether the high-redshift and radio-weak objects are included
in this optical sample, the overall levels of polarization observed are 
intermediate between those seen for X-ray and radio-selected
BL~Lac objects.

\end{abstract}


\section{Introduction}

Traditionally, BL~Lacertae objects have been found from radio and X-ray surveys 
\citep[see e.g.,][]{stickel91,stocke91}, or from comparisons between
source catalogs in these two wavelength regions \citep{muehleisen99}.
Past optical searches have had very little success
in finding these rare types of active galactic nuclei
\citep[AGN; see e.g.,][]{jannuzi93a}.
In addition, searches using a variety of methods have failed to
identify examples of radio-quiet BL~Lac objects in analogy to
radio-quiet QSOs \citep[e.g.,][]{impey82,borra84,stocke90,jannuzi93a}

It is easy to appreciate the difficulties encountered by optical
searches.
First, the main optical signatures of BL~Lac objects are not among
the properties that have been exploited to identify other types of AGN.
For example, their strong and rapid variations in flux and polarization have
been of little help since large-scale surveys based on variability that
have the required depth are
simply too difficult and time consuming given currently
available instrumentation.
Second, the lack of strong spectral lines in BL~Lac objects
denies the easiest method to spectroscopically identify members 
of the class.
This is particularly true for objects at the lower-luminosity end of the 
BL~Lac object distribution, because starlight from the host galaxy 
can help to mask the featureless synchrotron continuum, making broad-band
color selection very difficult.

Despite the inherent problems in optically searching for BL~Lac objects,
finding a sample of these AGN in a manner completely independent from
their radio and X-ray properties would be an important advance.
For instance, a purely optically-selected sample would explore the
range of properties exhibited by BL~Lac objects
with presumably different selection biases than those encountered in
X-ray and 
radio surveys.
In addition, deep optical surveys have the potential to provide
a more complete census of BL~Lac objects than
the relatively
shallow, large-area X-ray and radio surveys currently available.

In recent years, the
Sloan Digital Sky Survey \citep[SDSS;][]{york00} and the Two-Degree
Field QSO Redshift Survey \citep[2QZ;][]{boyle00}
have presented an opportunity to search for BL~Lac objects by exploiting one
of their defining characteristics: their featureless optical
continua.
Both surveys have been used to produce catalogs of
candidate BL~Lac objects by finding quasi-featureless spectra within their
huge spectroscopic databases.
\citet{londish02} describe the selection and properties of the 2QZ
BL~Lac object sample that contains over 50 candidates.
For the SDSS, \citet{collinge05} (C05) have used similar optical selection
criteria to identify 386 candidates.

An obvious way to confirm if these optical surveys are successful in 
finding true BL~Lac objects is to exploit another of the defining features
of this AGN class; namely, their high optical polarization.
In this paper, we present optical polarimetry of more than
40 BL~Lac candidates from the 
SDSS sample.
These measurements are a direct
test of the synchrotron
nature assumed to be responsible for the observed featureless continuum.
They also provide a basis for comparison with the polarization properties
of X-ray and radio-selected BL~Lac objects and offer an interesting new
sample for follow-up observations in other spectral regions.

\clearpage

\section{The Object Sample and Observations}

Objects were chosen for the optical polarization survey from the sample of
BL~Lac
candidates derived from the SDSS by C05.
Candidate selection by C05 is based on optical properties, most
notably a nearly featureless spectrum combined with proper motion and
broad-band color criteria to minimize Galactic contamination
by objects like DC white dwarfs.
The sample's 386 members are separated into ``probable''
and ``possible'' BL~Lac object categories by C05
depending on the likelihood that
a candidate is actually a Galactic star.
Forty-two of 156 probable BL~Lac candidates that are not already 
classified in the NASA Extragalactic Database (NED) as BL~Lac objects
were measured for optical polarization.
Although most of the objects in the C05 sample
that are already identified in NED
have not been observed for polarization, we exclude these objects from this
study to mitigate biases caused by the spectroscopic target selection
criteria used for the SDSS \citep[see e.g.,][]{ivezic02,anderson03}.
Specifically, 55 BL~Lac candidates were originally
targeted for spectroscopy because they
have counterparts detected by the Faint
Images of the Radio Sky at Twenty cm \citep[FIRST;][]{becker95} and/or
the {\sl ROSAT\/} All-Sky Survey \citep[RASS;][]{voges99}.
As a result,
the selection of these BL~Lac
candidates by C05 is obviously influenced
by the fact that spectra were obtained because of non-optical properties.
By ignoring the candidates previously identified in NED,
all of the objects found by other search techniques are excluded
from the polarization survey.
This criterion yields a sample completely independent of past
BL~Lac object searches, but
biases the polarization sample by selecting 
against the objects with the strongest radio and/or X-ray fluxes that can 
also be found in an unbiased optical survey.
In this case, 29 of 240 probable BL~Lac
candidates were excluded from the polarization 
survey even though they were chosen for SDSS spectroscopy based solely on their
optical properties.
The polarization survey targets are listed in Table~1 along
with their Galactic latitude ($b\/$),
SDSS $r\/$-band magnitude, estimated Galactic $r\/$-band extinction 
($A_r\/$), and
extinction-corrected optical spectral index, $\alpha_{\rm OPT}\/$, as listed or
derived from the SDSS database (C05).

Observations were made during 2005 May and
2006 October using the Steward Observatory 2.3~m Bok Telescope on
Kitt Peak, AZ and the SPOL spectropolarimeter \citep{schmidt92}.
Most objects were observed with SPOL in its imaging 
polarimetry configuration \citep[see e.g.,][]{smith02}.
In this mode, a mirror replaces the dispersive grating in the optical
train of the instrument and two $51\arcsec \times 51\arcsec$\/
images with complementary polarization are focused onto the thinned,
anti-reflection-coated $1200 \times 800$-pixel SITe CCD detector.
The CCD pixel size is 15~$\mu\/$m ($\sim0\farcs5$ on the sky).

The 2005 observations used a KPNO ``nearly Mould'' $R\/$ filter
(6000--7000~\AA).
A combination of Hoya HA-30 and Y-48 filters was selected for the imaging
polarimetry in 2006 October
to permit a broader spectral response of $\sim$4800--7000~\AA.
A $6\arcsec\/$ circular aperture was used in the reductions 
to determine the normalized linear polarization Stokes parameters, $q\/$ and
$u\/$.
The resulting degree of polarization ($P\/$) and polarization
position angle ($\theta\/$) are given in Table~1.
Unless otherwise noted, $P\/$ has been corrected for statistical
bias \citep{wardle74}, although this correction is very small
for most objects because
of the generally high signal-to-noise ratios (S/N)
of the measurements (the median $\sigma_P\/ = 0.32$\%). 

Spectropolarimetry was acquired for five of the brighter SDSS targets in the
sample on UT 2006 October~27.
The instrumental configuration and data reduction are similar to those
described in \citet{smith03}.
A slit width of $3\farcs0$ was employed along with spectral
extraction apertures of $10\arcsec\/$--$14\arcsec\/$.
The listed measurements of $P\/$ and $\theta\/$
in Table~1 for these objects are derived from averaging $q\/$ and
$u\/$ over the 5000--7000~\AA\ spectral range (500 pixels).
Both $P\/$ and $\theta\/$ show no significant variation
with wavelength across the entire spectrum for these objects.

We consider an object to be highly polarized if $P > 3$\%, in keeping
with past surveys of AGN \citep[see e.g.,][]{moore84}.
This criterion
has been effective at discriminating between sources with synchrotron
emission-dominated optical continua and those without. 
Also, this level of polarization is generally
sufficient to rule out a significant Galactic interstellar polarization
(ISP) contribution 
to the observed polarized flux for objects
that are more than $\sim$20\dgr\ away from the Galactic plane.
Indeed, the small estimated values of $A_r\/$ for the SDSS sample imply that
$P_{\rm ISP} \ll 3$\% for the candidates, 
even allowing for the empirically determined maximum ISP
for a given extinction,
$P_{\rm max} < 9.0E\/$($B-V\/$) = 3.27$A_r\/$ \citep{serkowski75}.
Polarimetry of objects included within the fields of several
BL~Lac candidates support the claim that ISP is not
significant. 
For field objects bright enough to yield polarization measurements of
high quality, i.e.,
$P/\sigma_P > 3$ and $\sigma_P < 1$\%, only those in the fields
of SDSS J083943.36+354001.5 (SDSS~J0839+3540; hereafter the IAU designation for
objects is shortened to SDSS~J{\it hhmm\/}$\pm${\it ddmm\/})
and SDSS~J2248$-$0036 show evidence for $P_{\rm ISP}\/$ approaching
2\%.
This high level of ISP is dwarfed
by the measured polarization of SDSS~J0839+3540.
In contrast, inclusion of the ISP along the sight line to
SDSS~J2248$-$0036 does not increase its polarization over the $P > 3$\%
threshold.

\section{Results}

Twenty-four of the 42 objects observed show optical polarization above 3\%,
verifying that a large number of candidates in the SDSS sample are indeed
BL~Lac objects.
The highest observed polarization is 22.6$\pm$0.1\% and there are 10
objects with $P > 10$\%.
This result proves that the approach used by C05 to identify
BL~Lac objects based primarily on their
optical spectra is generally
effective at finding new members of this class of AGN.
Beyond the basic finding that the SDSS sample is highly polarized,
the distribution of polarization within the sample
can be examined in various ways and
the results of this analysis are summarized in Table~2.

The sample can be split into two groups depending on the nature of the SDSS
spectra.
There are 15 objects in the polarization survey whose redshifts cannot
be determined
from the identification of
host galaxy absorption features.
These objects constitute the portion of the sample that shows the
highest polarization.
All 10 objects with $P > 10$\% have undetermined redshifts.
This pronounced selection effect is likely to be caused by the fact that
a featureless synchrotron continuum completely dominates
the optical spectra of these objects.
Any spectral features from the galaxy host are washed out at the
S/N level of the available spectroscopy, much like they are
for a large fraction of radio-selected BL~Lac objects that also
generally show high polarization (see \S4).

Only two of 15 objects with unknown $z\/$ fall into the low-polarization
category.
This is in marked contrast to those SDSS BL~Lac candidates with
measured redshifts.
More than half of the latter objects (16/27) are not observed to be highly
polarized, and there are no objects with $P > 10$\%.

Some of the discrepancy in the overall polarization levels between objects
with redshifts and those without
is caused by a larger contribution of the host galaxy
to the optical flux, at least for 
those objects with $z < 1$.
The dilution of synchrotron light by unpolarized starlight decreases
the observed polarization and this effect can be seen in the
imaging polarimetry for several SDSS objects.
Nine objects included in the polarization survey are classified as resolved
by C05 from the SDSS images.
Five of these meet the criterion for high polarization, but the 
maximum polarization observed for this group in a 6$\arcsec\/$ aperture
is $P = 7.1\pm0.3$\%.
Figure~1 displays the polarization as a function of observing aperture
for eight objects with host galaxies that are prominent enough
to be identified in the SDSS and easily seen in the polarimetry images.
For five objects, a clear trend with $P\/$ decreasing with increasing 
aperture
is apparent.
A similar analysis for highly polarized objects classified as point sources
by C05 reveals typical variations in polarization with aperture of
$\lesssim$0.05$P\/$.
The trend displayed in Figure~1 is in the expected sense, because the
measured polarization declines as more
unpolarized starlight is
included within the aperture.
The effect is most pronounced for SDSS~J2248$-$0036, however the nearly factor
of 2 increase in $P\/$ between a 6$\arcsec$ and  2$\arcsec$ aperture
is not enough for this candidate to meet the high-polarization criterion.
Likewise, correcting for host galaxy starlight does not elevate the
high-polarization resolved BL~Lac objects into the range of $P > 10$\%
that is populated exclusively by objects with unknown redshifts.
Note that Galactic ISP
can be ruled out as the sole polarizing mechanism for
low-polarization objects like SDSS~J2248$-$0036 because $P\/$ would
be constant with aperture.

The sample of objects with measured $z\/$ can be further subdivided into
$z < 1$ (18 objects) and  $z > 1$ (9 objects) groups.
Again, there is a large difference in
the distribution of polarization between
these two groupings.
High polarization is not observed in any of the high-redshift objects.
In fact, all of these objects are measured to have $P < 2$\% (see \S4).
The 11 highly polarized objects are found at $0.151 < z < 0.795$.

\citet{collinge05} identify 27 objects from their list of probable
BL~Lac candidates as being potentially radio-quiet.
Ten of these objects were observed for polarization.
Since this group consists mostly of objects with $z > 1$ (8 of 10 objects),
the overall levels of polarization measured are similar to the 
high-redshift sample.
That is, no objects were observed to be highly polarized.
In addition, only upper limits could be placed on $P\/$ for two
objects with unknown redshifts included
in this group: SDSS~J0201+0025 and SDSS~J1048+6203.
The 1-sigma upper limit on $P\/$ for SDSS~J0201+0025 is $\sim$5\%, so further
polarimetry is required to definitively determine if it is highly
polarized.

Finally, high polarization in the SDSS sample appears to be related
to the shape of the optical continuum as well as to whether discrete
spectral features are detected. 
For the entire polarization sample, we have corrected the SDSS $ugriz\/$
photometry for interstellar extinction and fit a power law to the 
resulting broad-band flux densities.
This is a crude approximation since 
a power law will not yield an acceptable fit if the host galaxy is an
important source of flux.
Also, fitting a power law over the entire optical spectrum will not be
adequate for high-$z\/$ objects as the Lyman~$\alpha\/$ forest and 
Lyman limit invade various filter bandpasses.

With these limitations in mind, we have calculated a spectral index,
$\alpha_{\rm OPT}\/$, where $F_\nu \propto \nu^{\alpha_{\rm OPT}}$.
The indices are listed in Table~1, and in the left panel of Figure~2
$P\/$ is plotted against $\alpha_{\rm OPT}\/$ for
all of the SDSS objects in the survey.
High-polarization objects are peaked strongly around $\alpha_{\rm OPT} = -1.5$.
All but three highly polarized objects fall within the range 
$-2 < \alpha_{\rm OPT} < -1$.
In contrast, objects with $z > 1$ span nearly the entire spectral
index range,
yet none are highly polarized.

In the right panel of Figure~2, $P\/$ is plotted against the ``goodness
of fit'' of a power law to the broad-band spectrum as measured
by the reduced $\chi^2$, $\chi_r^2$.
Higher polarization tends to be observed more often if the optical continuum 
is reasonably well fit by a simple power law.
A single power law is a poor fit to the continua of
most of the high-redshift objects mostly
owing to the Lyman~$\alpha\/$ forest and
Lyman limit spectral features
reducing flux in the blue bandpasses, and this is reflected in the 
very large $\chi_r^2$ for these objects.
The value of $\chi_r^2$ is listed for each object in Table~1.

\section{Discussion}

\subsection{Effectiveness of BL~Lac Object Selection from the SDSS}

The results of a polarization survey of 42 optically-selected
BL~Lac candidates from the SDSS show that a majority of
objects (24) are highly polarized and that their classification as
BL~Lac objects is confirmed.
The strategy employed by C05 of selecting quasi-featureless spectra from a deep
optical survey such as the SDSS and investigating proper motion
and broad-band color properties to minimize contamination by featureless
Galactic objects is largely validated by optical polarimetry.
The limiting magnitudes of the SDSS and 2QZ survey \citep[e.g.,][]{londish02}
can provide
a much deeper census of the BL~Lac object population than currently
available X-ray and radio surveys that until now have been the
most effective ways of finding these AGN.
The polarimetric verification of the SDSS sample is an important step
in showing that a purely optical search can be at least as efficient 
in this regard as surveys at X-ray and radio wavelengths.

There is, however, a relatively small subgroup within the probable
BL~Lac candidate list of C05 that is not polarimetrically confirmed. 
Namely, objects with redshifts $\gtrsim$1 selected from the SDSS and
objects that are necessarily much weaker radio sources than known BL~Lac
objects, given their
optical brightness,
have fundamentally different polarization properties from 
the rest of the sample.
All of these objects are measured to have very low polarizations.

BL~Lac objects are not highly polarized all of the time, but to have
none of the nine $z > 1$ candidates polarized $>3$\% in a synoptic
survey seriously challenges their classification as BL~Lac objects.
Among even the relatively optically sedate X-ray-selected
BL~Lac objects,
roughly 4 out of 10 will exhibit $P > 4$\% in a single-epoch survey
\citep{jannuzi94}.
If the continua of the SDSS high-$z\/$ candidates
are dominated by synchrotron emission, then
the magnetic fields within the emitting regions
must be unusually chaotic.

It might be argued that the discrepancy between the
general polarization levels of high-$z\/$
objects and the rest of the sample is caused by the dilution of the
synchrotron
light by an unpolarized continuum source that dominates the flux
in the ultraviolet.
A possible candidate is the Big Blue Bump (BBB) that dominates the 
UV-optical continuum of most strong broad emission-line AGN 
\citep[see e.g.,][]{malkan82,malkan83}.
The effect of the BBB \citep[plus the smaller ``3000~\AA\ bump'';][]{wills85}
on the synchrotron polarization spectrum has been demonstrated for
several optically violent variable quasars and 3C~273
\citep[see e.g.,][]{smith86,impey89,wills92,smith93b}.
In all of these cases, $P\/$ decreases into the blue as the BBB becomes
increasingly more important relative to the redder synchrotron spectrum.
BL~Lac objects, however, show neither this necessary polarimetric signature
\citep[][and references therein]{smith96},
nor the spectral flattening required
if the BBB is an important contributor to the UV flux \citep{impey88}.
In this regard,
none of the candidates show any indication of their spectra
flattening into the blue from the SDSS data, and
the spectropolarimetry available for
six of the SDSS objects is completely consistent with
past studies of BL~Lac objects by showing no evidence for
dilution by the BBB.

Possibly the most dramatic evidence that the blue bump (or any other plausible
diluting component)
is, in general, not a significant
source of ultraviolet flux in BL~Lac objects is provided by  
simultaneous UV/optical polarimetry.
These observations reveal that
$P\/$ is as high, or even higher in the UV than at
optical wavelengths down to a rest-frame wavelength of
1400~\AA\ \citep{allen93,smith93a}.
Roughly the same rest-frame spectral region is sampled by these past
UV measurements as in the imaging polarimetry of the
high-$z\/$ BL~Lac candidates.
The SDSS objects, therefore, must be quite different from known 
BL~Lac objects if dilution of the synchrotron light is responsible
for the very low observed polarizations.

The small number of high-redshift objects included in the SDSS sample
may be lower-$z\/$ analogs to SDSS J153259.96$-$003944.1 \citep{fan99},
a $z = 4.62$ object that defies easy classification within any of the
traditional AGN families.
Like the $z > 1$ objects observed for this study, SDSS~J1532$-$0039 does not 
exhibit emission lines, and \citet{fan99} measured its
polarization to be $< 2$\%.
Further observations will be necessary to determine whether these objects
constitute a new class of extragalactic object, fundamentally different
from other high-$z\/$ emission-line AGN, or are indeed BL~Lac objects 
requiring that the traditional definition for this class be radically
modified \citep[see also][]{londish04}.

A small number of probable BL~Lac candidates are identified
by C05
that could be weaker radio sources than
previously seen within this class of AGN.
Many of the 27 potential ``radio-quiet'' candidates are also members of the 
$z > 1$ group of objects
discussed above.
Polarimetry of 10 potentially radio-weak BL~Lac objects yields no evidence that
these objects are strongly polarized.
This result suggests that the SDSS 
has not uncovered a population 
of truly radio-quiet BL~Lac objects, consistent with earlier findings from
X-ray-selected samples \citep{stocke90}.

\subsection{Comparison with Other BL~Lac Object Samples}

Disregarding the potentially radio-weak objects, which comprise $<$12\%
of the total number of probable candidates, cross-correlation of the SDSS
BL~Lac candidates with available X-ray and
radio catalogs by C05 reveals that the sample is consistent
with the X-ray--optical--radio spectral energy distributions of
previously known BL~Lac objects.
Likewise, our polarimetry survey shows that the levels of optical
polarization observed for the SDSS sample are consistent with those
found in other BL~Lac object samples.
In Figure~3, the polarization of the
SDSS sample is compared to an X-ray-selected and
a radio-selected sample
of BL~Lac objects.
Thirty-seven BL~Lac objects observed 
by \citet{jannuzi93b} were chosen for the X-ray-selected comparison sample.
These objects were drawn from the 
{\it Einstein\/} Extended Medium Sensitivity Survey
\citep[EMSS;][]{gioia90,morris91} and from the {\it HEAO~1\/} A-2
survey \citep{maccacaro84}.
Since \citet{jannuzi93b} monitored the optical flux and polarization
of this sample over a period of $\sim$3 years, the first polarization
measurement of each object was used for comparison to the single-epoch
survey of the SDSS BL~Lac objects.
Many of these first measurements only set upper limits on $P\/$, and
therefore,
the distribution of polarization for the X-ray-selected BL~Lac objects in
Figure~3 is likely to be biased toward higher values at low $P\/$
simply because we treat the upper limits as true detections to 
avoid potentially larger biases that would occur by comparing a
synoptic survey with a true monitoring program.
The radio-selected sample is drawn from the 5~GHz 1~Jansky catalog
\citep[1~Jy;][]{kuhr81} and consists of 34 BL~Lac objects.
Many of these objects have been optically observed
for nearly four decades and the polarization distribution displayed
in Figure~3 is constructed from ``first observations'' from various sources
\citep{altschuler76,angel80,sitko85,smith87,kuhr90,impey90,ballard90}.

The pronounced difference in the polarization distributions between
X-ray and radio-selected
BL~Lac objects, first described by \citet{jannuzi94}, is clearly
apparent in Figure~3.
Nearly all ($\sim90$\%) of the radio-selected objects have $P > 3$\%, whereas
such levels of polarization are measured in only about half of the 
X-ray sample. 
Additionally, no X-ray BL~Lac object was observed to have $P > 10$\%, though
5 of the 37 objects exhibited $10{\rm \%} < P < 16$\% at some point during the
3-year long monitoring program \citep{jannuzi93b}.
In contrast, $\sim$50\% of the 1~Jy BL~Lac objects will, on average,
have $P > 10$\% at any given epoch.

We also plot
two probability curves for the SDSS candidate BL~Lac
objects in Figure~3.
One curve represents the entire polarization survey sample of 42 objects,
while the other omits the $z > 1$ objects as well as SDSS~J0201+0025
and SDSS~J1048+6203
since their classification as true BL~Lac objects is questionable
as discussed in \S4.1.
The two curves, therefore, represent the extremes for the 
cumulative distribution of optical polarization for the portion
of the SDSS spectroscopic sample not previously identified as BL~Lac
objects in NED (see \S2).
The fraction of SDSS objects that exhibit $P > 3$\% is intermediate between
the X-ray and radio samples whether or not the suspect low-polarization
candidates are included.
Also, the fraction of low-polarization objects in the full SDSS sample
is roughly similar to the fraction seen among the X-ray-selected sample.
The ``edited'' SDSS sample provides a
closer match in redshift  with the X-ray sample, but
a more definitive comparison between the low-polarization
tails of the SDSS and X-ray distributions requires better matching 
of the measurement uncertainties.
Typical values for $\sigma_P\/$ for the polarimetry of the 
EMSS and $HEAO~1\/$ A-2 BL~Lac objects \citep{jannuzi93b}
are factors of 2--4$\times\/$ larger
than those estimated from the imaging polarimetry of the SDSS targets.

\section{Summary and Future Work}

Optical polarimetry of 42 of 240 probable
BL~Lac candidates spectroscopically
selected from the SDSS verifies that this survey identifies true BL~Lac objects,
with over half of the measured objects showing $P > 3$\%. 
The level of polarization is seen to be related to the 
properties of the observed optical spectrum.
The highest polarization tends to be observed in objects
whose host galaxy features are undetected and have undetermined
redshifts at the S/N of the SDSS spectroscopy.
High optical polarization also tends to be observed for objects with
$-1 > \alpha_{OPT} > -1.7$ and continua that are reasonably well approximated
by a single power law.
Furthermore, the general polarization levels observed for the SDSS sample
are intermediate between the levels found for previously studied
X-ray and radio-selected BL~Lac objects.
The polarization survey does not confirm that any of the handful of
potentially radio-weak candidates are, in fact, BL~Lac objects.

The success of the SDSS in finding an optically-selected sample
of BL~Lac objects is a major advance in determining an accurate census
of this rare class of AGN, at least for $z \lesssim 1$.
Still to be determined is whether the SDSS has identified BL~Lac
objects at $z > 1$, or has, alternatively,
found members of a new class of high-redshift
AGN devoid of emission lines, but also not showing the usual properties
associated with a synchrotron continuum.
Near-infrared observations that measure the rest-frame
optical polarization of these objects can definitively
determine if a highly polarized
synchrotron spectrum is being diluted by unpolarized UV flux
that is not detected in low-redshift BL~Lac objects.
It would be extremely interesting to verify
the existence of such a strong, presumably non-relativistically beamed,
ultraviolet continuum source, especially since it does not
power a rich emission-line spectrum.

Variability studies may also hold some clues as to the 
nature of the high-$z\/$ objects. 
Indeed, strong variability, both in flux and polarization,
in all wavelength bands
is a defining property of BL~Lac objects and the investigation
of this aspect is important for understanding
the full range of behavior of the SDSS BL~Lac objects.
In particular, a central question is whether optically-selected
BL~Lac objects will show variability behavior consistent with
the wildly variable radio-selected sources, the more mildly
variable X-ray-selected objects, or be intermediate to 
these extreme examples as suggested by their measured 
polarization levels.
In this regard, future studies utilizing data from
projects like the Panoramic Survey Telescope \& Rapid Response System
(Pan-STARRS) and the Large Synoptic Survey Telescope (LSST) are likely to
provide the most definitive examination of the optical flux variability of
BL~Lac objects and other types of AGN.

\acknowledgments

This work was supported through 
National
Aeronautics and Space Administration contract 1256424.  We thank
Bill Wood and the Bok Telescope operations staff for the flawless
operation of the telescope.
We also thank the referee for comments that clarified some aspects
of this study.


\clearpage

\begin{deluxetable}{lrrrrrrrcrr}
\tablecolumns{11}
\tablewidth{0pc}
\tabletypesize{\small}
\tablecaption{SDSS BL~Lac Object Candidates}
\tablehead{
\colhead{SDSS~J} &
\colhead{$z\/$\tablenotemark{a}} &
\colhead{$b\/$~~(\dgr\/)} &
\colhead{$r\/$\tablenotemark{a}} &
\colhead{$A_r\/$\tablenotemark{b}} &
\colhead{$\alpha_{\rm OPT}\/$\tablenotemark{c}} &
\colhead{$\chi_r^2$\tablenotemark{c}} &
\colhead{UT Date} &
\colhead{Filter\tablenotemark{d}} &
\colhead{$P\/$~~(\%)} &
\colhead{$\theta\/$~~(\dgr\/)}}

\startdata

001736.91+145101.9 & 0.3026  & $-$47.2 & 17.86 & 0.13 & $-$1.35 & 0.5 & 2006--Oct--28 & $HY\/$ &  5.34$\pm$0.21 &  18.9$\pm$1.1 \\
002401.29+160233.9 & \nodata & $-$46.3 & 18.87 & 0.18 & $-$1.38 & 2.3 & 2006--Oct--29 & $HY\/$ & 17.90$\pm$0.59 & 153.0$\pm$0.9 \\
003808.50+001336.6 & \nodata & $-$62.5 & 19.30 & 0.06 & $-$1.40 & 1.8 & 2006--Oct--29 & $HY\/$ &  8.81$\pm$0.35 &  83.1$\pm$1.1 \\
010326.01+152624.8 & 0.2461  & $-$47.3 & 18.07 & 0.28 & $-$2.34 & 20 & 2006--Oct--28 & $HY\/$ &  1.09$\pm$0.12 &  59.8$\pm$3.1 \\
011452.77+132537.5 & \nodata & $-$49.0 & 17.03 & 0.08 & $-$1.13 & 0.9 & 2006--Oct--27 & $S\/$ & 12.09$\pm$0.11 &  22.7$\pm$0.3 \\
020137.66+002535.1 & \nodata & $-$57.7 & 19.54 & 0.08 & $-$1.19 & 12 & 2006--Oct--29 & $HY\/$ & $<$5.04 & \nodata \\
022048.46$-$084250.4 & 0.5252  & $-$62.0 & 18.27 & 0.06 & $-$1.04 & 0.7 & 2006--Oct--29 & $HY\/$ &  9.31$\pm$0.23 &  96.8$\pm$0.7 \\
031712.23$-$075850.4 & 2.6993  & $-$50.6 & 18.82 & 0.20 & $-$0.68 & 3.5 & 2006--Oct--28 & $HY\/$ &  0.71$\pm$0.36 &  14.1$\pm$15.4 \\
032343.62$-$011146.1 & \nodata & $-$45.2 & 16.81 & 0.22 & $-$0.76 & 0.5 & 2006--Oct--27 & $S\/$ & 14.83$\pm$0.10 &  50.4$\pm$0.2 \\
074054.60+322601.0 & 0.9460  & $+$23.9 & 18.67 & 0.13 & $-$0.88 & 1.0 & 2006--Oct--27 & $S\/$ &  2.88$\pm$0.37 & 100.8$\pm$3.7 \\
074242.93+301836.2 & 0.4948  & $+$23.6 & 19.48 & 0.14 & $-$1.98 & 5.3 & 2006--Oct--30 & $HY\/$ &  6.87$\pm$0.48 &  54.1$\pm$2.0 \\
081609.58+491004.5 & \nodata & $+$33.7 & 17.93 & 0.15 & $-$1.58 & 2.1 & 2006--Oct--27 & $S\/$ &  7.98$\pm$0.25 & 171.0$\pm$0.9 \\
081751.01+324340.5 & 0.7950  & $+$31.5 & 18.53 & 0.11 & $-$1.17 & 0.9 & 2006--Oct--30 & $HY\/$ &  6.01$\pm$0.28 & 117.0$\pm$1.3 \\
083458.19+440338.2 & \nodata & $+$36.6 & 17.84 & 0.07 & $-$1.51 & 5.6 & 2006--Oct--27 & $S\/$ & 22.62$\pm$0.11 &   4.3$\pm$0.1 \\
083918.75+361856.1 & 0.3343  & $+$36.5 & 19.60 & 0.11 & $-$2.01 & 7.2 & 2006--Oct--30 & $HY\/$ &  3.55$\pm$0.44 &  44.8$\pm$3.5 \\
083943.36+354001.5 & \nodata & $+$36.5 & 18.77 & 0.11 & $-$1.31 & 1.0 & 2006--Oct--30 & $HY\/$ & 10.36$\pm$0.32 &  31.9$\pm$0.9 \\
084056.06+372105.8 & \nodata & $+$37.0 & 19.23 & 0.10 & $-$1.49 & 4.0 & 2006--Oct--28 & $HY\/$ & 11.94$\pm$0.44 & 138.5$\pm$1.0 \\
084411.67+531250.7 & \nodata & $+$38.1 & 18.32 & 0.07 & $-$1.26 & 5.6 & 2006--Oct--30 & $HY\/$ & 14.43$\pm$0.27 &  10.1$\pm$0.5 \\
085135.93+552834.5 & \nodata & $+$38.9 & 18.17 & 0.10 & $-$1.30 & 1.6 & 2006--Oct--30 & $HY\/$ & 10.03$\pm$0.13 &  39.3$\pm$0.4 \\
104833.57+620305.0 & \nodata & $+$49.6 & 19.85 & 0.03 & $-$1.08 & 0.5 & 2005--May--13 & $R\/$ & $<$2.33 & \nodata \\
114153.35+021924.4 & 3.5979  & $+$60.1 & 18.61 & 0.06 & $-$1.59 & 51 & 2005--May--13 & $R\/$ &  0.90$\pm$0.64 &  58.3$\pm$21.6 \\
121221.56+534128.0 & 3.1900  & $+$62.5 & 18.63 & 0.05 & $-$1.44 & 98 & 2005--May--14 & $R\/$ &  1.32$\pm$0.58 &  67.1$\pm$13.0 \\
123132.38+013814.0 & 3.2300  & $+$64.1 & 18.75 & 0.05 & $-$1.15 & 25 & 2005--May--13 & $R\/$ &  1.69$\pm$0.64 &  58.5$\pm$11.1 \\
123743.09+630144.9 & 3.5347  & $+$54.0 & 19.00 & 0.03 & $-$2.18 & 49 & 2005--May--14 & $R\/$ &  1.21$\pm$0.78 & 102.8$\pm$19.7 \\
142505.61+035336.2 & 2.2476  & $+$57.8 & 18.72 & 0.08 & $-$1.08 & 15 & 2005--May--13 & $R\/$ & $<$0.93 & \nodata \\
154515.78+003235.2 & 1.0114  & $+$40.6 & 18.82 & 0.27 & $-$0.47 & 9.6 & 2005--May--13 & $R\/$ &  1.05$\pm$0.69 &   7.3$\pm$20.0 \\
165808.33+615001.9 & 0.3742  & $+$37.0 & 18.48 & 0.13 & $-$1.95 & 8.2 & 2006--Oct--29 & $HY\/$ &  1.02$\pm$0.24 &  43.8$\pm$6.7 \\
170124.64+395437.1 & 0.5071  & $+$37.4 & 16.88 & 0.06 & $-$1.60 & 5.3 & 2006--Oct--28 & $HY\/$ &  9.08$\pm$0.14 &   8.5$\pm$0.4 \\
171445.55+303628.0 & 0.8500  & $+$33.1 & 19.11 & 0.13 & $-$0.29 & 1.4 & 2005--May--13 & $R\/$ & $<$0.90 & \nodata \\
205456.85+001537.8 & 0.1508  & $-$27.0 & 18.04 & 0.22 & $-$1.82 & 21 & 2006--Oct--29 & $HY\/$ &  3.36$\pm$0.15 & 134.9$\pm$1.3 \\
205528.20$-$002117.2 & \nodata & $-$27.5 & 17.96 & 0.26 & $-$0.53 & 0.7 & 2006--Oct--28 & $HY\/$ &  4.06$\pm$0.18 & 100.8$\pm$1.3 \\
211603.23$-$010828.4 & 0.3052  & $-$32.3 & 17.99 & 0.16 & $-$1.27 & 1.8 & 2006--Oct--28 & $HY\/$ &  5.11$\pm$0.11 &   4.4$\pm$0.6 \\
211611.89$-$062830.4 & 0.2916  & $-$35.0 & 18.85 & 0.36 & $-$1.71 & 5.3 & 2006--Oct--29 & $HY\/$ &  7.05$\pm$0.32 & 173.5$\pm$1.3 \\
224749.55+134248.2 & 1.1746  & $-$39.3 & 18.27 & 0.14 & $-$0.60 & 17 & 2006--Oct--28 & $HY\/$ &  0.75$\pm$0.22 &   0.1$\pm$8.0 \\
224819.44$-$003641.6 & 0.2123  & $-$50.3 & 18.77 & 0.26 & $-$2.37 & 10 & 2006--Oct--29 & $HY\/$ &  1.37$\pm$0.19 & 152.8$\pm$4.0 \\
224949.85+130844.7 & 0.6334  & $-$40.1 & 19.10 & 0.16 & $-$1.09 & 3.5 & 2006--Oct--28 & $HY\/$ &  2.88$\pm$0.47 & 174.0$\pm$4.6 \\
225232.18+124510.9 & 0.4966  & $-$40.8 & 18.82 & 0.21 & $-$1.53 & 5.2 & 2006--Oct--28 & $HY\/$ &  5.36$\pm$0.31 &  27.2$\pm$1.7 \\
225354.24+140436.9 & 0.3269  & $-$39.9 & 18.50 & 0.15 & $-$1.51 & 1.0 & 2006--Oct--28 & $HY\/$ &  2.19$\pm$0.14 & 106.5$\pm$1.9 \\
225624.27+130541.7 & \nodata & $-$41.0 & 18.57 & 0.12 & $-$1.40 & 1.4 & 2006--Oct--28 & $HY\/$ & 14.55$\pm$0.47 & 178.0$\pm$0.9 \\
231952.83$-$011626.8 & 0.2835  & $-$56.1 & 18.53 & 0.10 & $-$2.34 & 29 & 2006--Oct--29 & $HY\/$ &  3.82$\pm$0.23 & 123.1$\pm$1.7 \\
232428.43+144324.4 & 1.4100  & $-$43.1 & 18.78 & 0.13 & $-$0.62 & 21 & 2006--Oct--29 & $HY\/$ &  0.90$\pm$0.23 & 132.0$\pm$7.1 \\
233453.84+143214.8 & \nodata & $-$44.4 & 18.22 & 0.33 & $-$1.55 & 2.6 & 2006--Oct--28 & $HY\/$ & 16.46$\pm$0.77 & 177.1$\pm$1.3 \\

\enddata

\tablenotetext{a}{Data are from the SDSS database (C05).}
\tablenotetext{b}{The estimate of the $r\/$-band Galactic extinction in
magnitudes (C05).}
\tablenotetext{c}{The extinction-corrected optical spectral index 
and the reduced $\chi^2$ of a power-law fit to
the SDSS $ugriz\/$ photometry.}
\tablenotetext{d}{$HY =$ Hoya HA-30 + Y-48 filter; $S =$ spectropolarimetric
observation.}

\end{deluxetable}


\begin{deluxetable}{lcccccc}
\tablecolumns{7}
\tablewidth{0pc}
\tabletypesize{\small}
\tablecaption{Polarization of Optically-Selected BL~Lac Object Candidates}
\tablehead{
\colhead{Sample}  &
\colhead{$N\/$} & \colhead{$N\/$($P > 20$\%)} &
\colhead{$N\/$($P > 10$\%)} & \colhead{$N\/$($P > 3$\%)} &
\colhead{Median $P\/$ (\%)} &
\colhead{Range in $P\/$ (\%)} }
\startdata

All objects\tablenotemark{a}: & 42 & 1 & 10 & 24 & 4.6 & 0.7--22.6 \\
 & & & & & & \\
Objects with unknown $z\/$: & 15 & 1 & 10 & 13 & 11.9 & 2.3--22.6 \\
Objects with known $z\/$: & 27 & 0 & 0 & 11 & 2.2 & 0.7--9.3 \\
~~~~~Objects with $z < 1$: & 18 & 0 & 0 & 11 & 3.7 & 0.9--9.3 \\
~~~~~Objects with $z > 1$: & 9 & 0 & 0 & 0 & 0.9 & 0.7--1.7 \\
 & & & & & & \\
Resolved objects: & 9 & 0 & 0 & 5 & 3.4 & 1.0--7.1 \\
Potential radio-weak objects\tablenotemark{b}: & 10 & 0 & 0 & (1) & 1.0 & 0.7--$< 5.1$ \\

\enddata

\tablenotetext{a}{The four objects that have only upper limits for their
optical polarization are not considered to be highly polarized.}
\tablenotetext{b}{SDSS~J0201+0025 is the only object that could
conceivably be polarized above 3\% since the 1-sigma upper limit for
its polarization is $\sim 5$\%.}

\end{deluxetable}

\clearpage

\begin{figure}[t]
\figurenum{1}
\vspace{-40mm}
\hspace{25mm}
\includegraphics[scale=0.6,keepaspectratio=true]{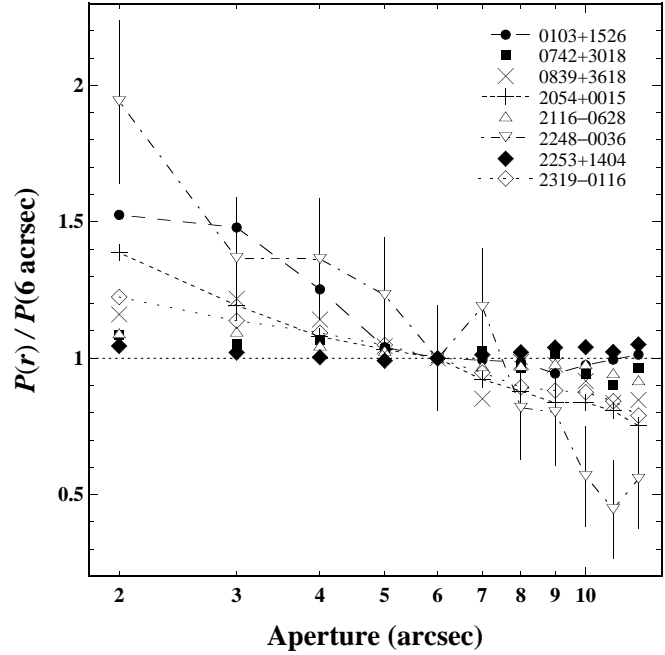}
\vspace{-25mm}
\caption{Polarization as a function of aperture for eight
BL~Lac candidates that are resolved in the SDSS images.
The measurements are normalized to the polarization observed in a
6$\arcsec$ aperture.
For most objects plotted, the trend is for $P\/$ to decrease as more
starlight from the galaxy host is included within the measurement aperture.
Representative error bars are shown for SDSS~J2248$-$0036.}
\label{fig1}
\end{figure}


\begin{figure}
\figurenum{2}
\vspace{-80mm}
\hspace{-10mm}
\includegraphics[scale=1.0,keepaspectratio=true]{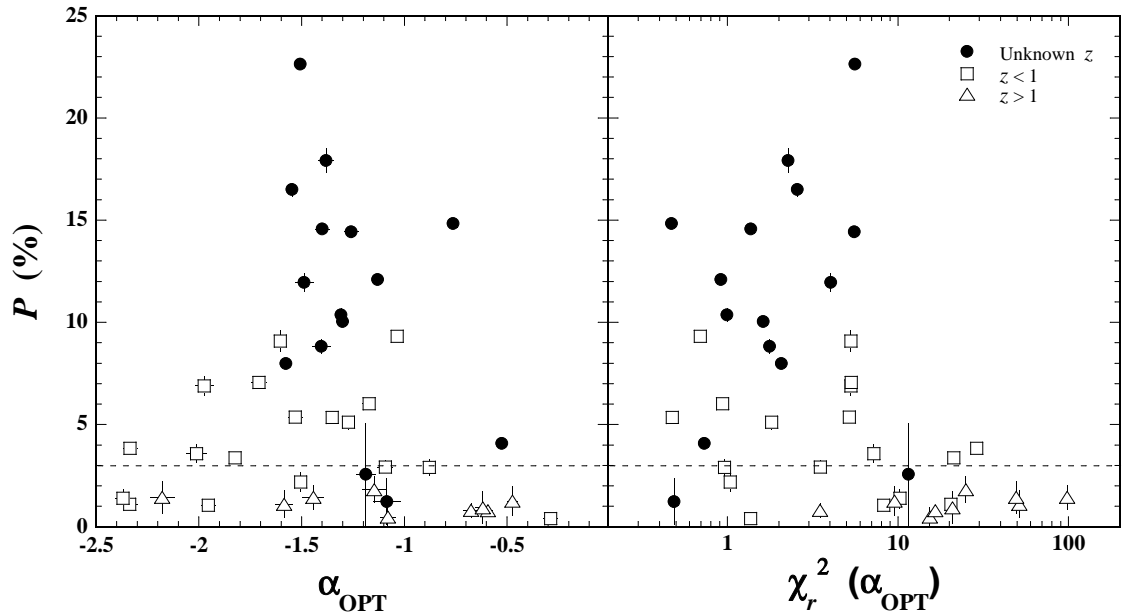}
\vspace{-95mm}
\caption{The observed polarization plotted against the optical
spectral index (left panel) and the reduced $\chi^2$ (right panel) for a
single power-law fit to the Galactic extinction-corrected, broad-band
SDSS photometry (see text).
In this figure, $P\/$ has not been corrected for statistical bias.
Objects with unknown redshift are represented by {\it filled\/}
circles.
Objects with measured redshifts ({\it open\/} symbols) are divided into
$z < 1$ ({\it squares\/}) and $z > 1$ ({\it triangles\/}) groups.
The {\it dashed\/} line at $P = 3$\% is the adopted division between
high- and low-polarization objects.}
\label{fig2}
\end{figure}

\begin{figure}
\figurenum{3}
\plotone{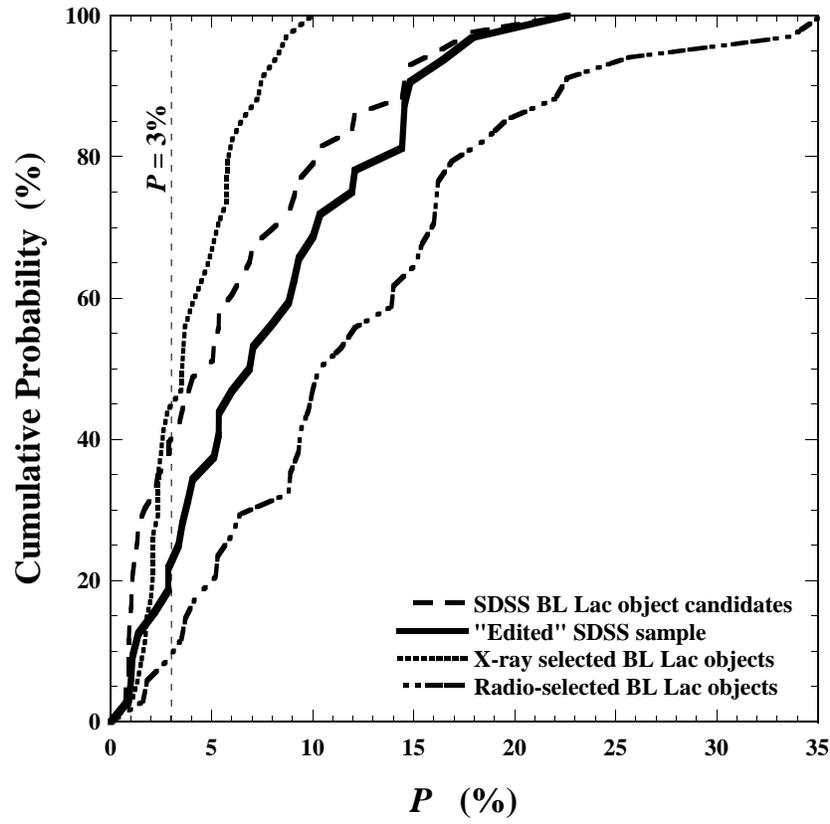}
\vspace{-30mm}
\caption{The cumulative distribution of optical
polarization for various samples of BL~Lac objects.
Two distributions are shown for the optically-selected SDSS BL~Lac objects.
All 42 objects of the polarization survey are represented by the 
{\it dashed\/} curve.
The {\it solid\/} 
curve represents the ``edited'' SDSS sample
that excludes all $z > 1$ members
as well as two additional low-polarization objects
that are radio weak
compared to known BL~Lac objects (\S4.1).
Nearly 80\% of the edited SDSS sample is highly polarized ($P > 3$\%).
See \S4.2 for a description of the membership of the comparison
X-ray and radio-selected samples.}
\label{fig3}
\end{figure}

\end{document}